\documentclass[tikz,conference]{IEEEtran}
\IEEEoverridecommandlockouts
\usepackage{amsmath,graphicx,cite}
\usepackage{amssymb}
\usepackage{graphicx}
\usepackage{color}
\usepackage{subcaption}

\usepackage[a4paper, top=1.9cm, bottom=2.54cm, left=1.65cm, right=1.65cm]{geometry}

\title{Deterministic Modeling of Dynamic ISAC Channels in RF Digital Twin Environments}

\author{
\IEEEauthorblockN{César Montaner, Saúl Fenollosa, Andres Ortega, Hugo Beltrán, Narcis Cardona}
\IEEEauthorblockA{\textit{iTEAM Research Institute, Universitat Politècnica de València} \\
cemoncar@iteam.upv.es, sjfenarg@upv.edu.es, alortort@iteam.upv.es, hbelsan@upv.edu.es, ncardona@iteam.upv.es}
}

\begin{document}

\maketitle

\begin{abstract}

This paper introduces a methodology to calibrate Radio-Frequency Digital Twins (RF-DTs) for Integrated Sensing and Communication (ISAC) in dynamic wireless environments. The approach leverages high-resolution ray tracing in combination with wideband channel sounding to ensure consistency between simulated and measured propagation. The methodology is validated in urban scenarios featuring both mono-static and bi-static configurations, as well as moving user platforms and vehicles. Results show that the calibrated RF-DT reproduces key propagation effects, including multipath evolution, dynamic scatterers, and Doppler-induced signatures, with close agreement to measurements. These findings confirm that accurate geometry, material modeling, antenna patterns, and diffuse scattering are essential for realistic high-frequency ISAC simulation. By bridging the gap between simulation and measurement, the proposed calibration framework provides a scalable tool for developing and evaluating ISAC algorithms in complex, time-varying environments envisioned for 6G.  

\end{abstract}

\begin{IEEEkeywords}
Digital Twin, RF-DT, Dynamic ISAC, Channel Sounding, Ray-Tracing, 6G Wireless Systems 
\end{IEEEkeywords}

\section{Introduction}
Integrated Sensing and Communication (ISAC) has emerged as a key paradigm for next-generation 6G systems, enabling wireless infrastructure to jointly support high-rate data transmission and environmental sensing~\cite{Lu2024ISAC}. Beyond spectrum reuse, ISAC pursues unified hardware and waveform design to deliver timely situational awareness for control, automation, and safety-critical services. In this context, accurate channel models are essential, as they determine the fidelity of sensing/communication co-design and the reliability of data-driven algorithms. Ray-tracing (RT) and radio-frequency digital twins (RF-DTs) have therefore gained traction for reproducing multipath propagation with high physical consistency~\cite{zhang2024digital}. Crucially, an RF-DT can generate and curate large, labeled, and scenario-specific datasets under controlled interventions, thereby supplying the data needed to train and stress-test learning-based ISAC pipelines while supporting continual updates as environments evolve.

RT-based models have traditionally emphasized specular propagation in static scenes. This abstraction is insufficient for ISAC, which must learn from dynamic and variable environments where non-specular energy and motion carry discriminative information about objects and events. In practice, incorporating a diffuse–scattering abstraction is required to account for power that is not captured by a purely specular model at millimeter-wave/sub-terahertz bands, while explicit mobility modeling is needed to reflect target- and platform-induced time variation. Experimental studies report that diffuse components can be a non-negligible share of received power in urban settings~\cite{Charb2020}, motivating their inclusion alongside mobility within RT-based simulators~\cite{Pascual,Bakirtzis2020,Yaman,Sheikh}. Efficient implementations, such as directive/diffuse scattering in Sionna RT, make this feasible at scale~\cite{Hoydis2023}.

\begin{figure}[t!]
  \centering
  \begin{subfigure}[b]{0.24\textwidth}
    \includegraphics[width=\linewidth]{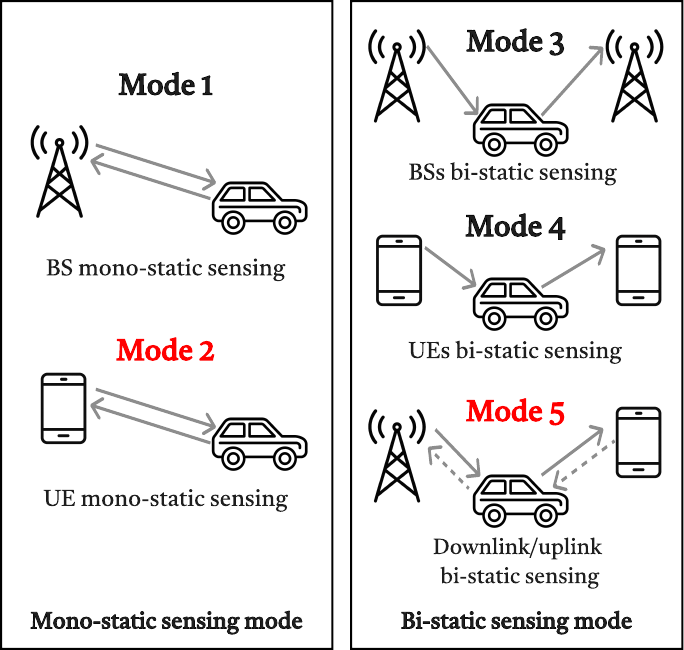}
    \caption{}
    \label{fig:isac_modes}
  \end{subfigure}\hfill
  \begin{subfigure}[b]{0.224\textwidth}
    \includegraphics[width=\linewidth]{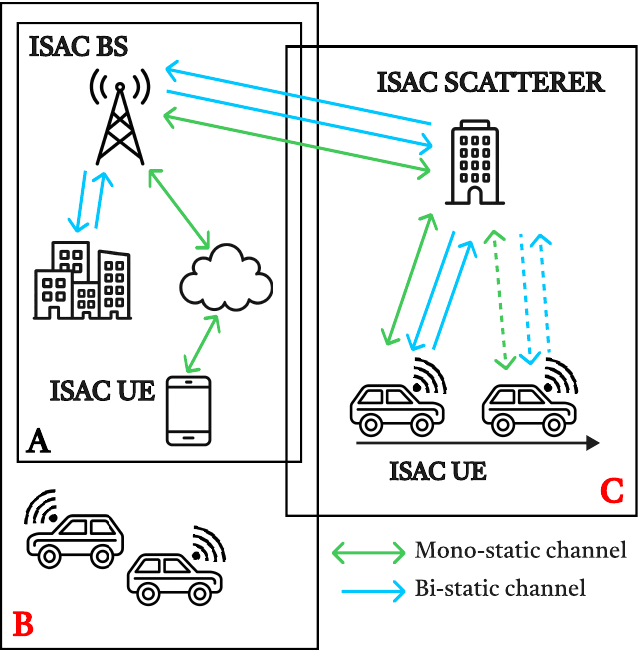}
    \caption{}
    \label{fig:isac_scenarios}
  \end{subfigure}
  \caption{Definition of ISAC Sensing Modes (a) and Channel Scenarios (b)}
  \label{fig:isac_taxonomy}
\end{figure}

The present study adopts the ISAC mode/scenario taxonomy of~\cite{yang2023isac}, observed in Fig.~\ref{fig:isac_taxonomy}. Sensing modes are grouped into mono-static (transmitter and receiver co-located; Modes~1--2) and bi-static (transmitter and receiver separated; Modes~3--5). Scenarios characterize motion: (A) static transceivers and static channel elements, (B) static transceivers with moving elements, and (C) moving transceivers. Our focus is on realistic, dynamic operation, i.e., Scenarios~B and~C, with particular attention to User Equipment (UE) mono-static (Mode~2) and downlink/uplink Base Station (BS)–UE bi-static (Mode~5) sensing.

This paper instantiates and validates a calibrated RT-based RF-DT for ISAC at 77–81 GHz tailored to Scenarios~B/C and Modes~2/5 in Urban Microcell (UMi) settings. Given a geometrically consistent 3D representation of the scene and time-aligned trajectories for all moving entities, we: (i) perform synchronized wideband measurements with a channel sounder over representative vehicle-mobility episodes; (ii) instantiate an RT simulation of the same episodes using the measured transceiver states and tracked object kinematics; and (iii) include material-parameterized directive/diffuse scattering to account for non-specular paths. This enables a rigorous, phase-coherent comparison between simulation and ground-truth measurements over delay and Doppler, thereby assessing the extent to which a calibrated RF-DT can reproduce dynamic ISAC channels.

The remainder of the paper is organized as follows. Section~II details the methodology, covering the measurement campaign, scene geometry/kinematics acquisition, and RT configuration. Section~III reports validation results against measurements across Modes~2/5 in Scenarios~B/C. Section~IV discusses implications for ISAC system design and data-driven sensing, and Section~V concludes.


\section{Methodology}
\label{sec:methodology}
This section specifies a single, end-to-end procedure to instantiate and validate a calibrated RT-based RF-DT at 77--81\,GHz for ISAC Modes~2/5 under dynamic UMi Scenarios~B/C. The content is organized as follows: (i) measurement campaign and hardware; (ii) scenario design; (iii) scene/kinematics acquisition; (iv) RT configuration with directive/diffuse scattering; (v) dynamic channel representation and delay--Doppler extraction; and (vi) alignment and comparison protocol consistent with the figures reported in the Results. The objective is a like-for-like comparison between measured and simulated delay--Doppler signatures under realistic motion and non-specular propagation.

\subsection{Measurement Campaign and Hardware}
A frequency-modulated continuous-wave (FMCW) sensing platform performs wideband E-band channel sounding using mutual-interference processing for bi-static characterization~\cite{Cardona2023Radar, Castilla2024DualISAC}. Each unit integrates three TX and four RX antennas; operation uses 4\,GHz bandwidth centered at 79\,GHz. Key radio specifications and experiment parameters are summarized in Tables~\ref{tab:sensor_specs} and~\ref{tab:sensor_configs}.

\begin{table}[t!]
    \centering
    \caption{Sensor Specifications}
    \label{tab:sensor_specs}
    \begin{tabular}{|l|c|}
        \hline
        \textbf{Parameter} & \textbf{Value} \\ \hline
        Center Frequency & 79 GHz \\
        Bandwidth & 4 GHz \\
        TX / RX Antennas & 3 / 4 \\
        Peak Gain (per antenna) & 10.5 dBi \\
        H / V Beamwidth (3 dB) & $\pm 28^\circ$ / $\pm 14^\circ$ @ 78 GHz \\
        TX Power & 12 dBm \\
        Noise Figure & 14 dB (76--77 GHz), 15 dB (77--81 GHz) \\
        \hline
    \end{tabular}
\end{table}

\begin{table}[t!]
    \centering
    \caption{Measurement Configuration Parameters}
    \label{tab:sensor_configs}
    \begin{tabular}{|l|c|}
        \hline
        \textbf{Parameter} & \textbf{Value} \\ \hline
        Ramp Duration ($T_\text{chirp}$) & 112.86 $\mu$s \\
        Idle Time ($T_\text{idle}$) & 13 $\mu$s \\
        Chirp Slope ($S$) & 35.44 MHz/$\mu$s \\
        I/Q Sampling Rate ($f_s$) & 18.75 MHz \\
        Max Range (Mono/Bi-static) ($R_\text{max}$) & 79.4 / 158.8 m \\
        \hline
    \end{tabular}
\end{table}

The array layout affords azimuthal coverage suited to UMi links, while the 4\,GHz bandwidth ensures fine range resolution for separating nearby multipath components (MPCs) and enabling short-window Doppler estimation.

\subsection{Scenario Design (UMi LOS)}
Experiments were conducted on the Universitat Polit\`ecnica de Val\`encia campus (UMi, LOS). A BS node was placed at the fourth floor of Building~6D, illuminating the courtyard between buildings. Two dynamic episodes were executed to realize Scenarios~B/C and Modes~2/5:

\textit{Configuration B (Scenario B; Modes 2/5).}
BS and UE are static; a Nissan Micra follows an approximately linear trajectory parallel to a facade. The BS points to the UE for a stable BS--UE link (Mode~5), and the UE boresight is oriented toward the vehicle path to enhance mono-static returns (Mode~2).

\textit{Configuration C (Scenario C; Modes 2/5).}
The UE is roof-mounted on a moving KIA Xceed whose trajectory is roughly perpendicular to the BS boresight. Platform motion excites rich near-field interactions (e.g., bike racks, lamp posts) and strong far reflections from buildings.

\begin{figure}[t!]
  \centering
  \begin{subfigure}[b]{0.48\textwidth}
    \includegraphics[width=\linewidth]{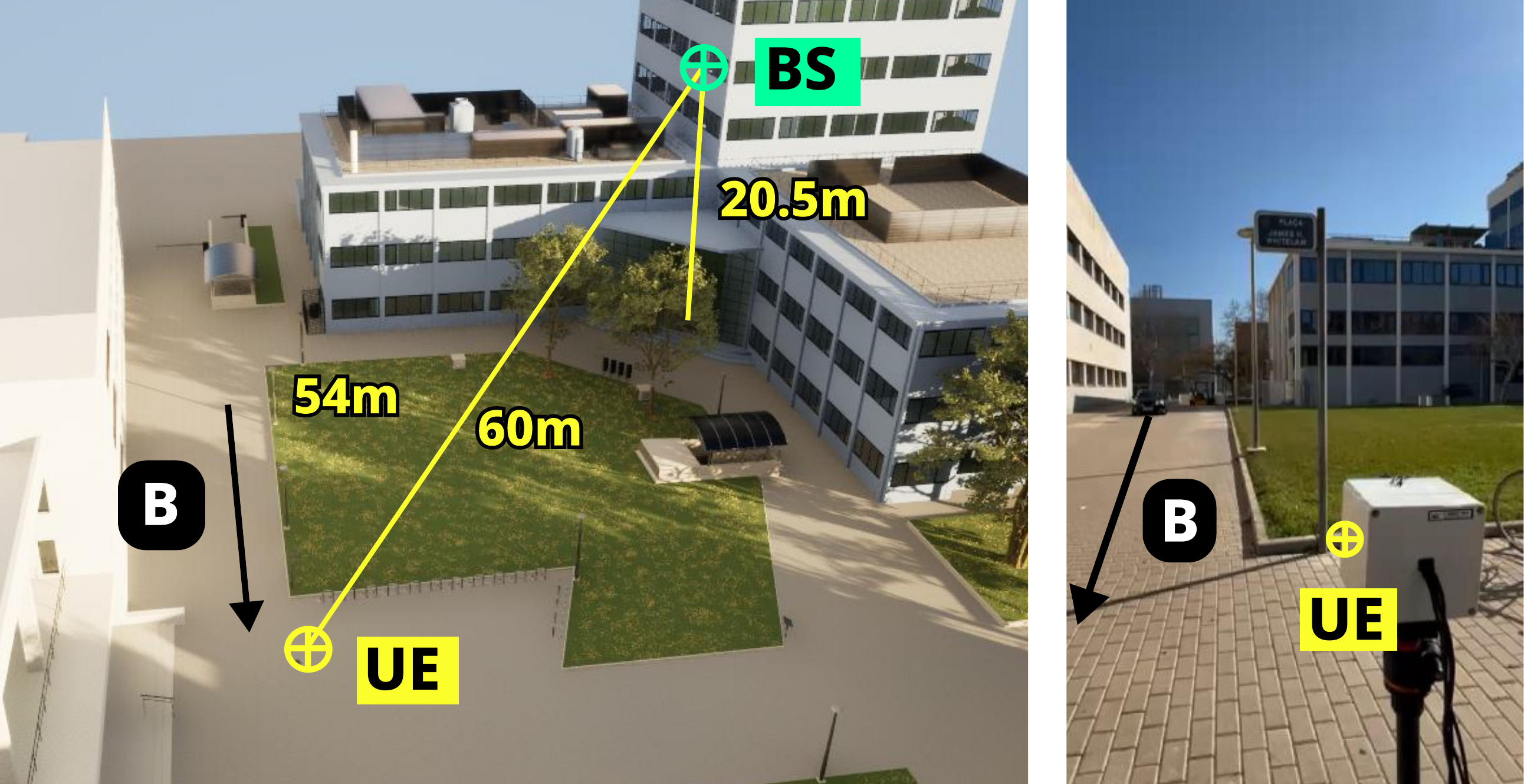}
    \caption{}
    \label{fig:ScenarioB}
  \end{subfigure}\hfill
  \begin{subfigure}[b]{0.48\textwidth}
    \includegraphics[width=\linewidth]{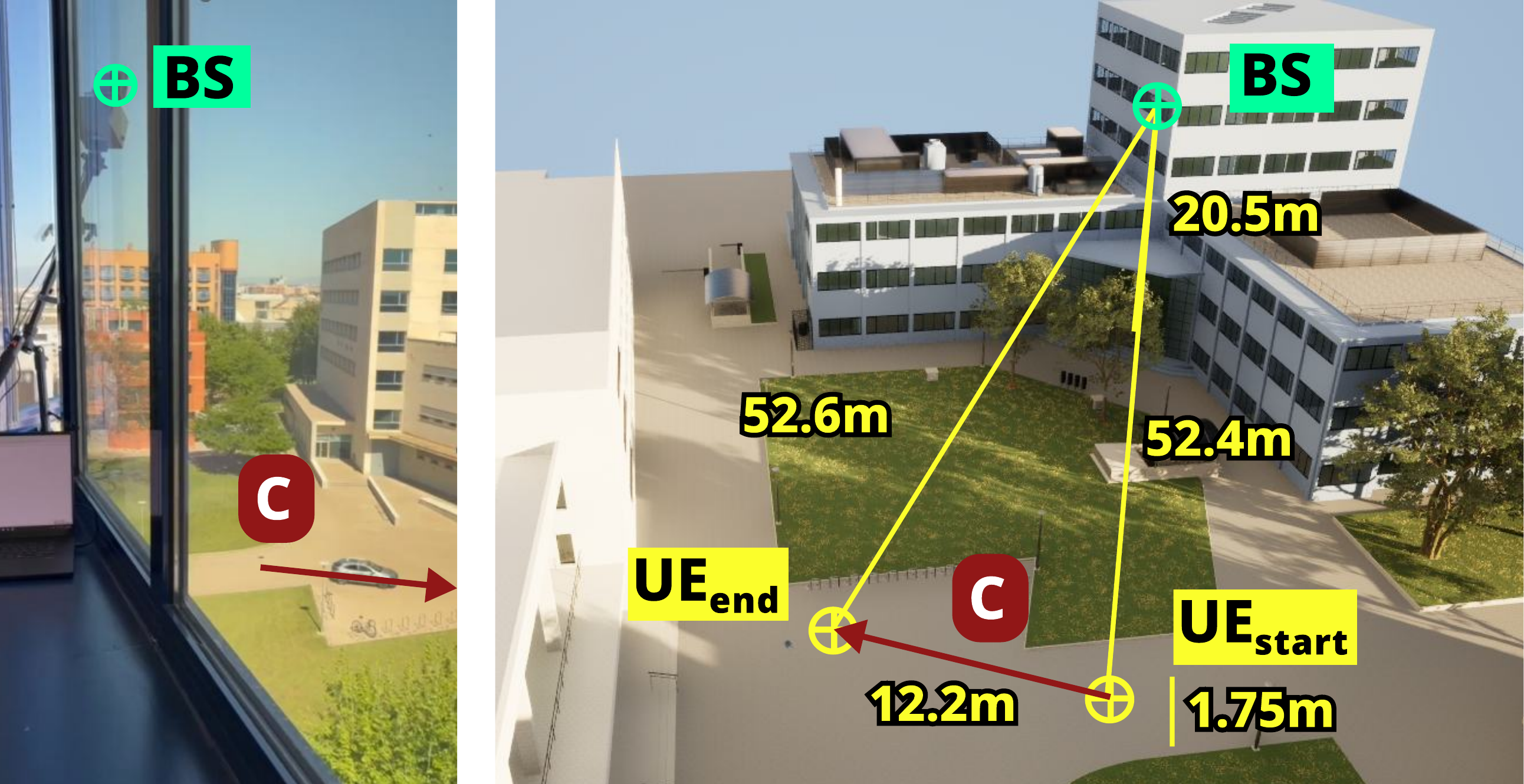}
    \caption{}
    \label{fig:ScenarioC}
  \end{subfigure}
  \caption{Measurement episodes in UMi LOS: (a) Scenario-type~B with static BS/UE and moving target; (b) Scenario-type~C with moving UE platform.}
  \label{fig:scenariosB}
\end{figure}

\subsection{Scene and Kinematics Acquisition}
A metrically consistent $3$D model of the courtyard and surrounding facades is used to instantiate the RF-DT. Absolute transceiver positions, boresights, and timing are logged per chirp. Vehicle trajectories are time-stamped and smoothed to enforce kinematic consistency, yielding time-aligned states for all moving entities required by the dynamic RT engine.

\subsection{RT Configuration with Directive/Diffuse Scattering}
\begin{figure}[t!]
  \centering
  \begin{subfigure}[b]{0.14\textwidth}
    \includegraphics[width=\linewidth]{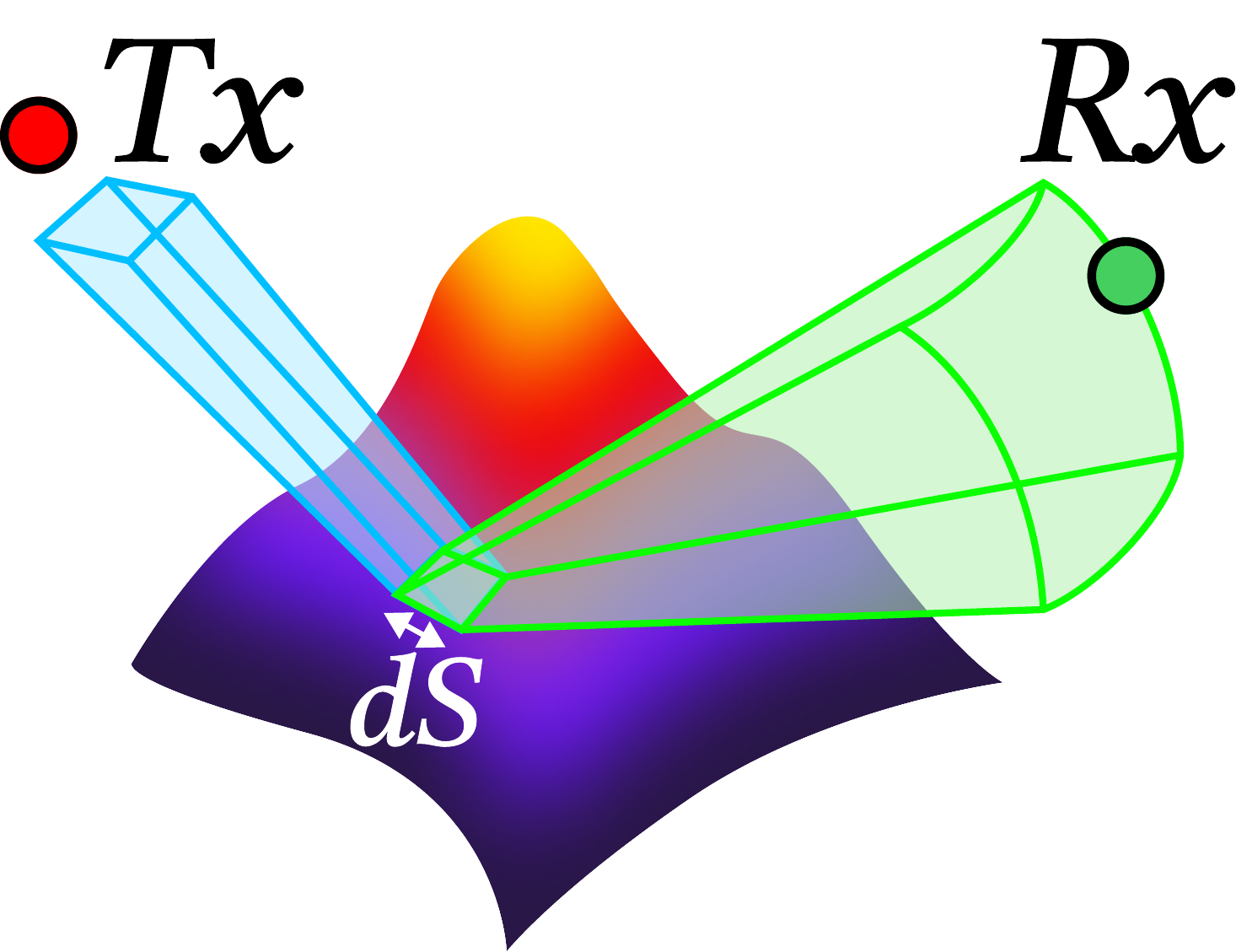}
    \caption{}
    \label{fig:continuous_surface}
  \end{subfigure}\hfill
  \begin{subfigure}[b]{0.14\textwidth}
    \includegraphics[width=\linewidth]{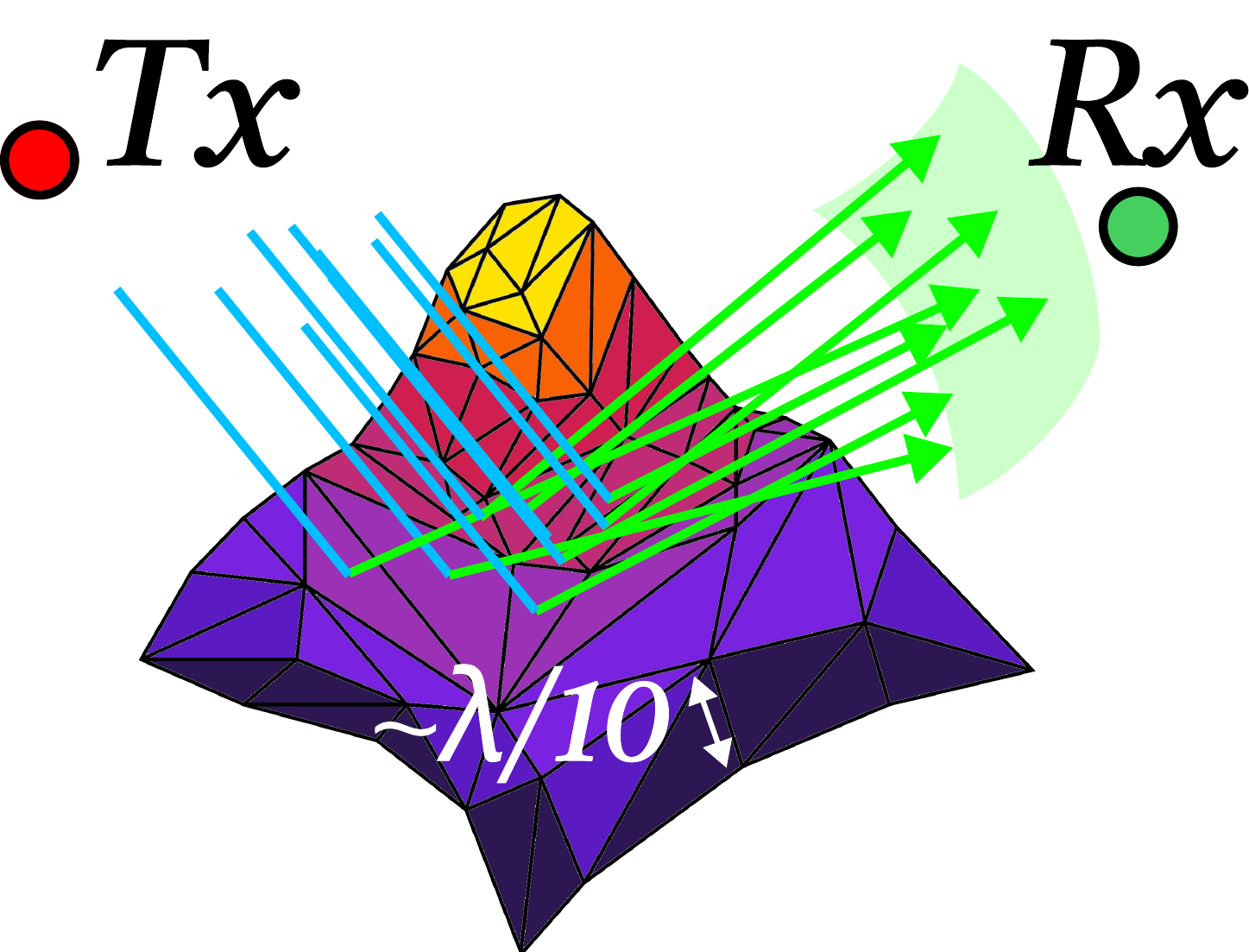}
    \caption{}
    \label{fig:sampled_surface}
  \end{subfigure}
  \begin{subfigure}[b]{0.17\textwidth}
    \includegraphics[width=\linewidth]{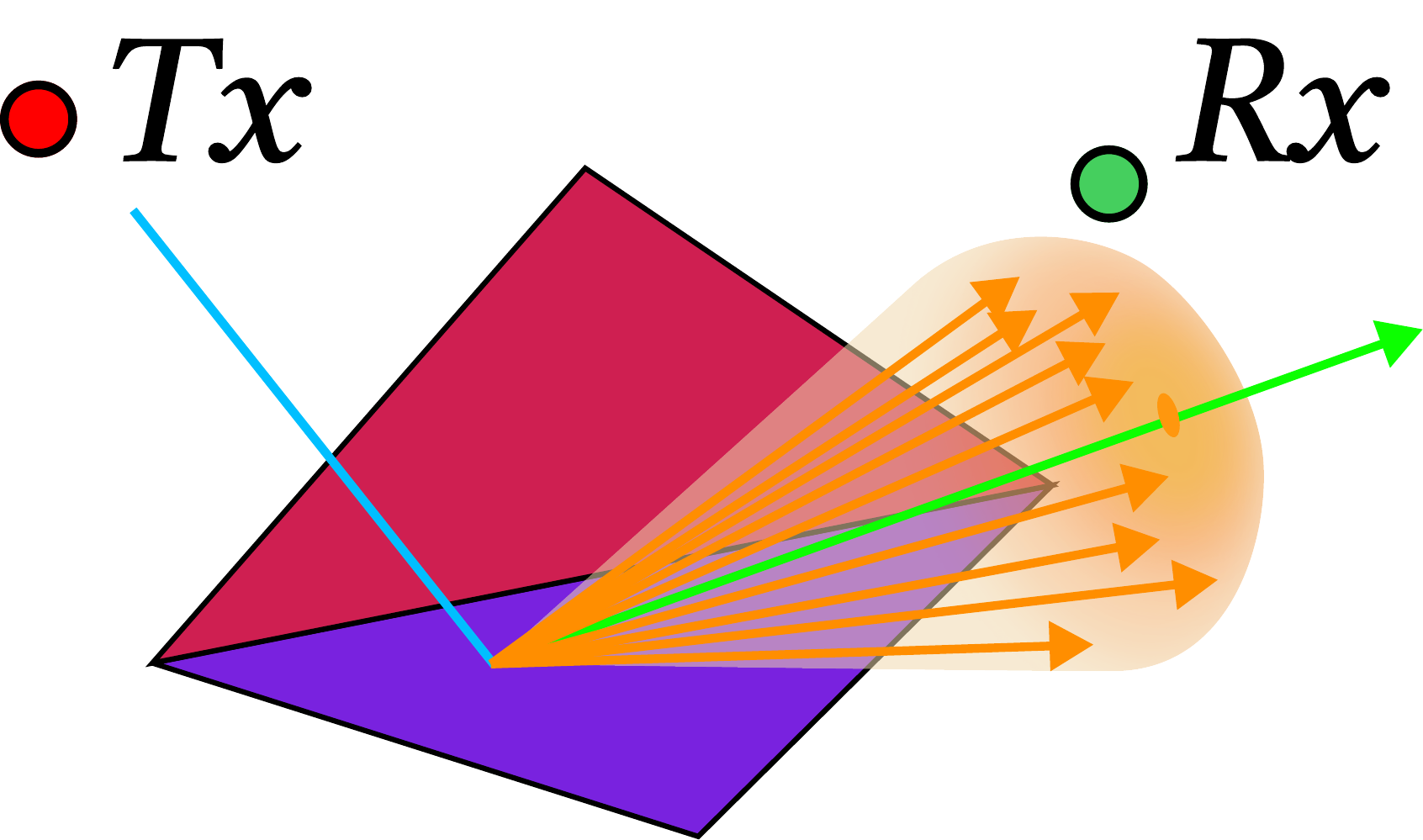}
    \caption{}
    \label{fig:simplified_surface}
  \end{subfigure}
  \caption{Scattering abstraction: (a) physical scattering on rough surfaces; (b) micro-geometry discretization; (c) specular reflection plus directive/diffuse lobe on macro facets.}
  \label{fig:scattering_abstraction}
\end{figure}

Purely specular RT on low-polygon meshes underestimates power at mmWave/sub-THz due to electromagnetic roughness; direct meshing at $\mathcal{O}(\lambda/10)$ is infeasible at E-band~\cite{taflove2004fdtd}. Based on a three-level scattering abstraction, summarized in Fig.~\ref{fig:scattering_abstraction}, a diffuse scattering abstraction is therefore adopted in Sionna RT~\cite{Hoydis2023}, splitting incident power into specular and diffuse components with conservation
\begin{equation}
    R^2 + S^2 = 1,
    \label{eq:scat_coef}
\end{equation}
where $R$ is reflection reduction and $S$ the scattering coefficient. The diffuse lobe is directive around the specular direction,
\begin{equation}
    f_\text{s}(\hat{\mathbf{k}}_s) \propto 
    \left(\frac{1+\hat{\mathbf{k}}_r^\mathsf{T}\hat{\mathbf{k}}_s}{2}\right)^{\alpha_s(S)},
    \label{eq:directive_scat}
\end{equation}
with $\alpha_s$ controlling lobe sharpness. Material-dependent $S$ values, informed by empirical data~\cite{Guo2025Meas}, are assigned to typical urban surfaces (glass, concrete, metal, brick). The interaction order captures dominant specular plus first-order diffuse paths while avoiding path explosion~\cite{Hoydis2023}. Motion updates delays, angles, and Doppler at the chirp rate for Scenarios~B/C.

\subsection{Dynamic Channel Representation and Delay--Doppler Extraction}
To accurately model the time-varying characteristics of sensing channels in dynamic scenarios, the channel impulse response (CIR) is expressed as \eqref{eq:cir}. Each multipath component $n$ is characterized by a time-varying complex amplitude $a_n(t)$, propagation delay $\tau_n(t)$, and a Doppler shift $\nu_n(t)$. This model is valid both for real measurements and RT simulations, provided that the dynamic nature of the environment (e.g., moving scatterers or transceivers) is properly captured.

\begin{equation}
    h(t,\tau) = \sum_{n=1}^{N}a_{n}(t)\delta\left( \tau - \tau_{n}(t) \right)e^{j2\pi\nu_{n}(t)t}
    \label{eq:cir}
\end{equation}

To obtain the Doppler-delay spectrum, $H(\nu, \tau)$, the Short-Time Fourier Transform (STFT) of the CIR is computed with respect to time over a window of duration $T_w$, as defined in \eqref{eq:stft}.

\begin{equation}
    H(\nu, \tau) = \int_{t_0}^{t_0 + T_w} h(t, \tau) e^{-j2\pi \nu t} dt
    \label{eq:stft}
\end{equation}

The key to simplifying the model is the assumption that in the short integration window $T_w$, the parameters for each path are approximately constant, where $a_n(t) \approx a_n$, $\tau_n(t) \approx \tau_n$, and $\nu_n(t) \approx \nu_n$. This leads to a locally stationary approximation of the channel impulse response. Applying these approximations, \eqref{eq:fft_finite} is obtained by substituting \eqref{eq:cir} into \eqref{eq:stft}.

\begin{align}
    H(\nu, \tau) &\approx \int_{t_0}^{t_0 + T_w} \left( \sum_{n=1}^{N} a_n \delta(\tau - \tau_n) e^{j2\pi\nu_n t} \right) e^{-j2\pi \nu t} dt \nonumber \\
    &\approx \sum_{n=1}^{N} \int_{t_0}^{t_0 + T_w} a_n \delta(\tau - \tau_n) e^{j2\pi\nu_n t} e^{-j2\pi \nu t} dt \nonumber \\
    &\approx \sum_{n=1}^{N} a_n \delta(\tau - \tau_n) \int_{t_0}^{t_0 + T_w} e^{j2\pi(\nu_n - \nu)t} dt
    \label{eq:fft_finite}
\end{align}

The remaining integral is the Fourier transform of a complex exponential over a finite window. The magnitude of this integral is proportional to the $sinc$ function:
\begin{equation}
     \left| \int_{t_0}^{t_0 + T_w} e^{j2\pi(\nu_n - \nu)t} dt \right| \propto T_w \cdot \text{sinc}\left( (\nu_n - \nu) T_w \right) 
     \label{eq:doppler_mag}
\end{equation}

This leads to the final expression for the Doppler-delay spectrum \eqref{eq:doppler}, where $e^{j\phi_n}$ is a complex phase term resulting from the integration bounds.

\begin{equation}
    H(\nu, \tau) \approx \sum_{n=1}^{N} \underbrace{ \left( a_n e^{j\phi_n} \right) }_{\substack{\text{Complex} \\ \text{Amplitude}}} \cdot \underbrace{ \delta(\tau - \tau_n) }_{\substack{\text{Path Delay} \\ \text{Impulse}}} \cdot \underbrace{ T_w \cdot \text{sinc}\left( (\nu_n - \nu) T_w \right) }_{\substack{\text{Doppler} \\ \text{Profile}}}
    \label{eq:doppler}
\end{equation}

The structure of \eqref{eq:doppler} explains the output of a dynamic sensor-based system, justifying the behavior observed in real-world measurements:
\begin{itemize}
    \item \textbf{Path Delay Impulse}: The $\delta(\tau - \tau_n)$ term ensures that each multipath component appears as a sharp peak at its specific propagation delay $\tau_n$.
    \item \textbf{Doppler Profile}: The $\text{sinc}( (\nu_n - \nu) T_w )$ term describes the shape of the component in the Doppler domain. The $sinc$ function is maximized when $\nu = \nu_n$, meaning that each component's energy is concentrated in a peak centered at its true Doppler shift $\nu_n$.
\end{itemize}

Static objects do not produce a Doppler shift ($\nu_n = 0$ Hz), resulting in a central peak at zero frequency in the Doppler spectrum. In contrast, moving objects induce non-zero Doppler shifts ($\nu_n \neq 0$), which appear as distinct peaks offset from the center, directly indicating their relative velocities. This behavior enables comparison between measured Doppler-delay spectra (via FFT) and RT-predicted Doppler shifts, facilitating phase-aware validation of the RF Digital Twin in dynamic ISAC scenarios.

\subsection{Alignment and Comparison Protocol}
Processing mirrors the pipeline used to produce the figures in the Results, so that measurements and simulations are treated identically. Temporal alignment is performed at the chirp level using hardware timestamps, and delay alignment follows the measured FMCW timing model, enabling direct comparison of CIR delays without additional phase calibration. Doppler information is extracted from short sequences of $N\!=\!128$ consecutive chirps, defining
\begin{equation}
    T_w \;=\; N\,\big(T_{\text{chirp}} + T_{\text{idle}}\big),
    \label{eq:t_window}
\end{equation}
with $T_{\text{chirp}}$ and $T_{\text{idle}}$ as in Table~\ref{tab:sensor_configs}. Within each window, motion manifests as the phase evolution $e^{j2\pi \nu_n t}$ of~\eqref{eq:doppler}. The same windowing and DFT parameters are applied to both datasets. On the simulation side, the RT engine provides path delays and powers synchronized to the chirp grid; Doppler shifts $\nu_n$ are computed from instantaneous radial velocities of the corresponding rays and used to modulate each path across the $N$ chirps, yielding delay--Doppler spectra consistent with~\eqref{eq:doppler}. Comparisons are then performed on power--delay profiles over time and on delay--Doppler maps computed on matched windows, emphasizing agreement of dominant components in delay, the stability of the $\nu\!\approx\!0$ ridge from static scatterers, and the location of peaks associated with moving elements in Configurations~B/C. Figures report side-by-side measured and simulated outputs with identical axes, windowing, and color scales.

\section{Results}

This section evaluates the calibrated RF-DT against measurements across Scenarios~B/C and Modes~2/5. Results are organized by scenario, with emphasis on the agreement of power--delay structure, Doppler dynamics, and their relevance for ISAC operation. The analysis uses the aligned processing chain described in Section~\ref{sec:methodology}, ensuring that measurements and simulations are directly comparable.

\subsection{Scenario B: Static Transceivers with Moving Vehicle}
\begin{figure}[t!]
  \centering
  \begin{subfigure}[b]{0.242\textwidth}
    \includegraphics[width=\linewidth]{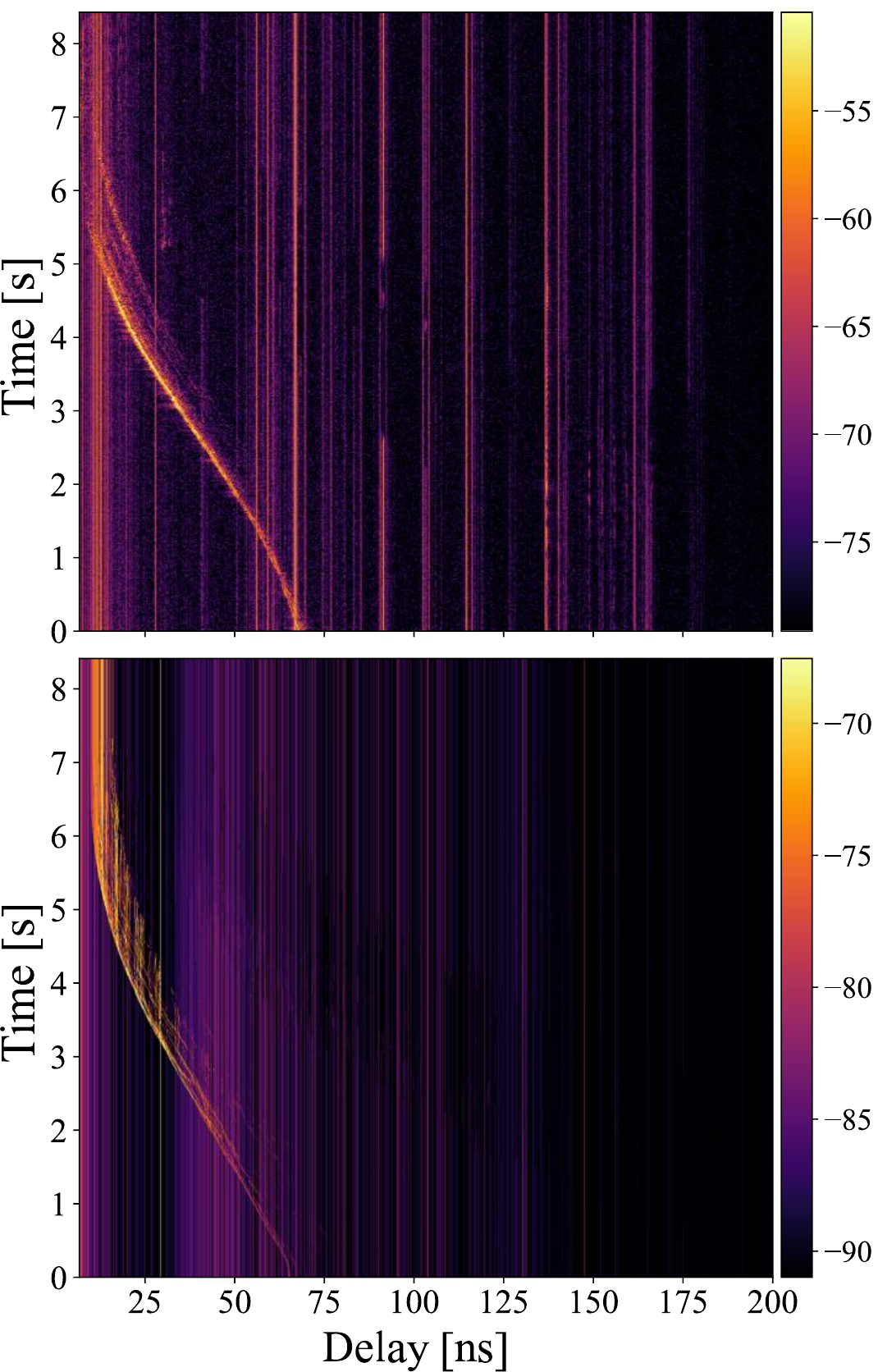}
    \caption{}
    \label{fig:B1_mono}
  \end{subfigure}
    \begin{subfigure}[b]{0.238\textwidth}
    \includegraphics[width=\linewidth]{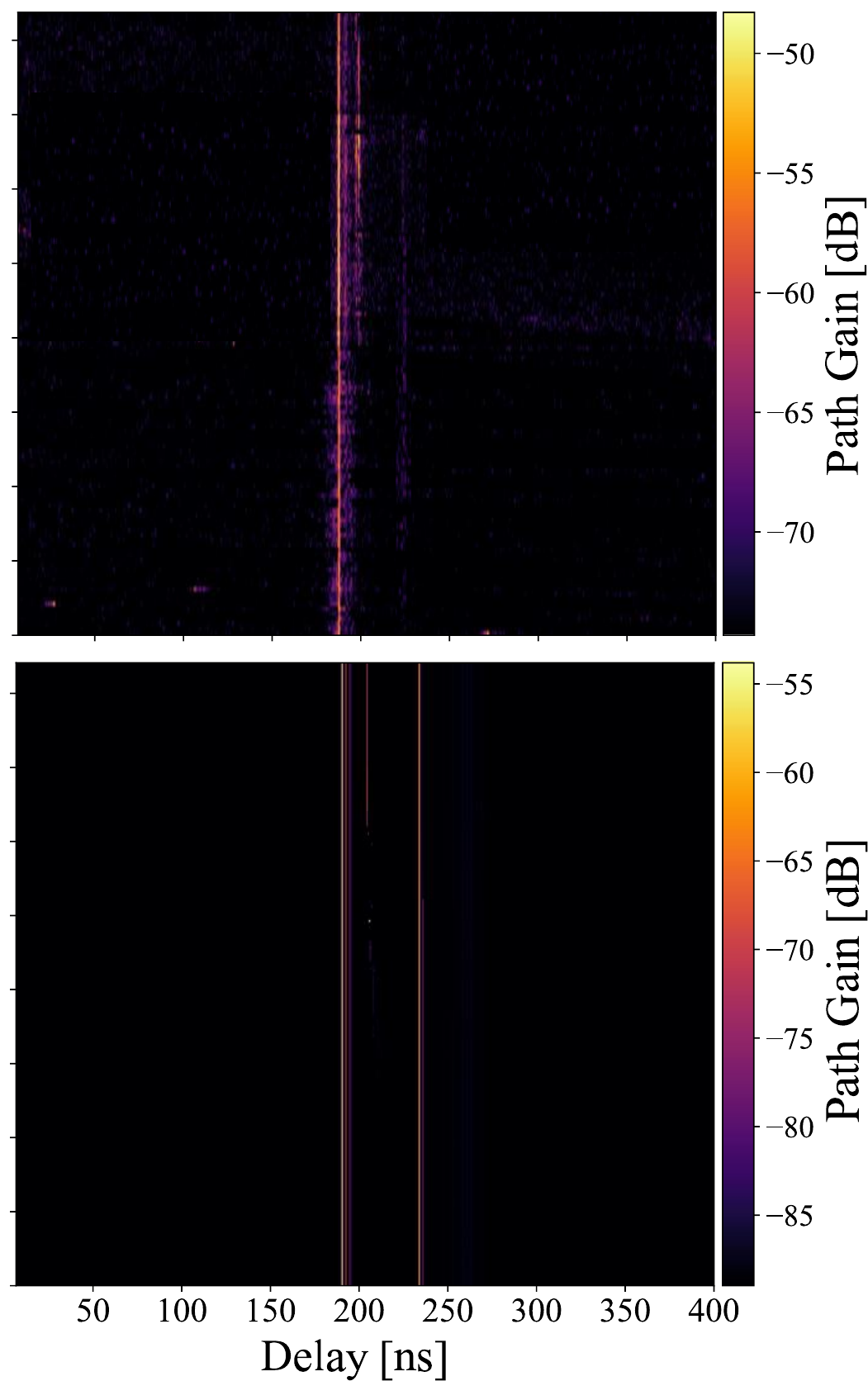}
    \caption{}
    \label{fig:B1_bi}
  \end{subfigure}
  \caption{Scenario~B: PDPs for mono-static (a) and bi-static (b). Top: measurement; bottom: simulation.}
\end{figure}

Scenario~B considers static BS/UE nodes and a moving car. In the mono-static case (Fig.~\ref{fig:B1_mono}), the approaching vehicle creates a clear time-varying path that shifts in delay as the car moves. This trajectory is consistently observed in both measurement and simulation, indicating that the calibrated RF-DT can correctly reproduce moving-target signatures.  

In the bi-static setup (Fig.~\ref{fig:B1_bi}), distinct scatterers contribute: the LOS path, the vehicle, and reflections from building facades. Each appears as a separate MPC cluster, visible in both datasets. The RT faithfully reproduces their relative delays and temporal evolution, which is critical for ISAC applications where distinguishing between static and dynamic reflectors underpins localization and tracking tasks.

This confirms that the RF-DT can reproduce mixed environments with both static and moving scatterers, ensuring that vehicle-induced signatures remain distinguishable against strong static multipath.

\subsection{Scenario C: Moving UE Platform}
\begin{figure}[t!]
  \centering
    \begin{subfigure}[b]{0.242\textwidth}
    \includegraphics[width=\linewidth]{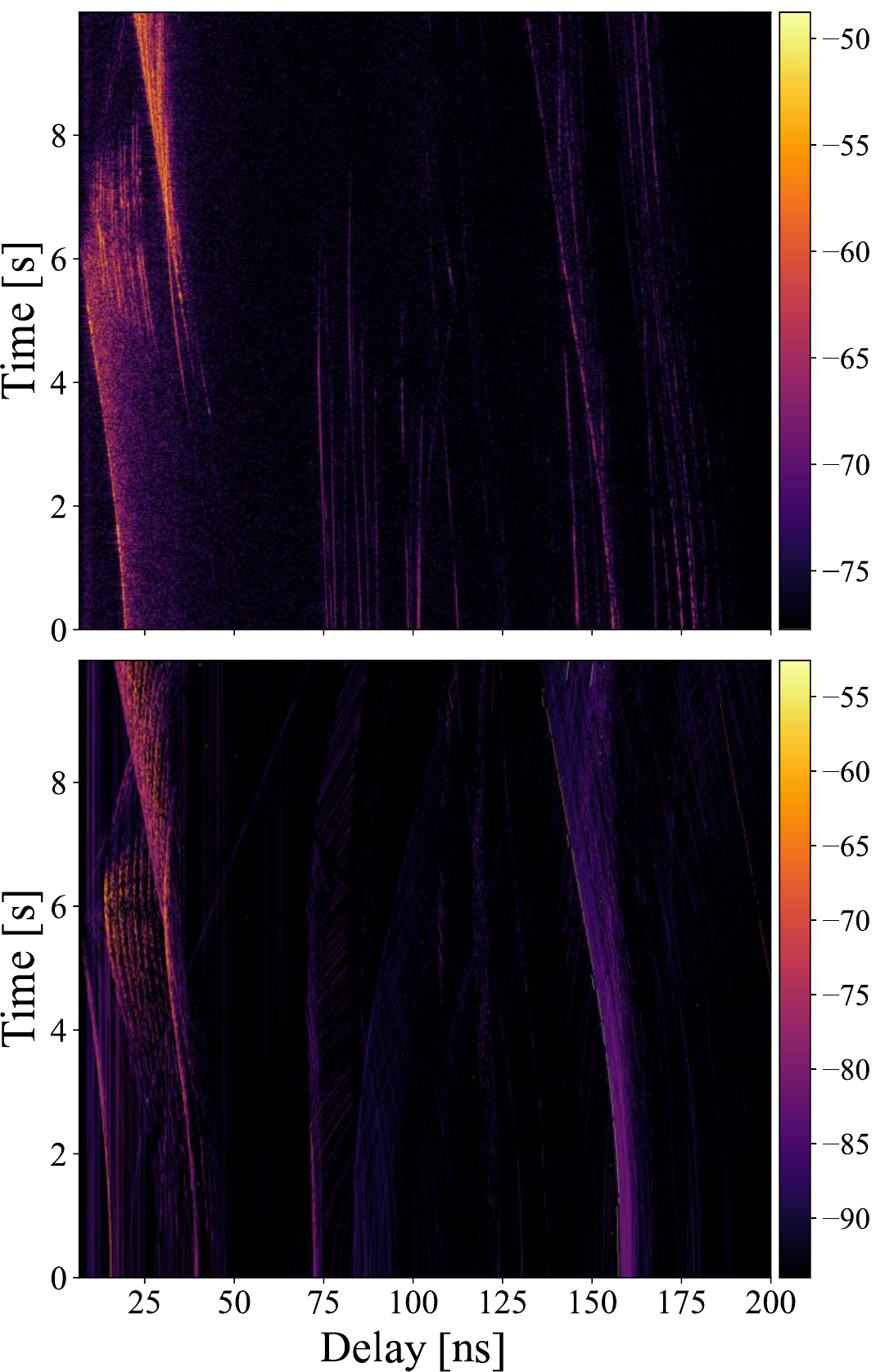}
    \caption{}
    \label{fig:C6_mono}
  \end{subfigure}
  \begin{subfigure}[b]{0.238\textwidth}
    \includegraphics[width=\linewidth]{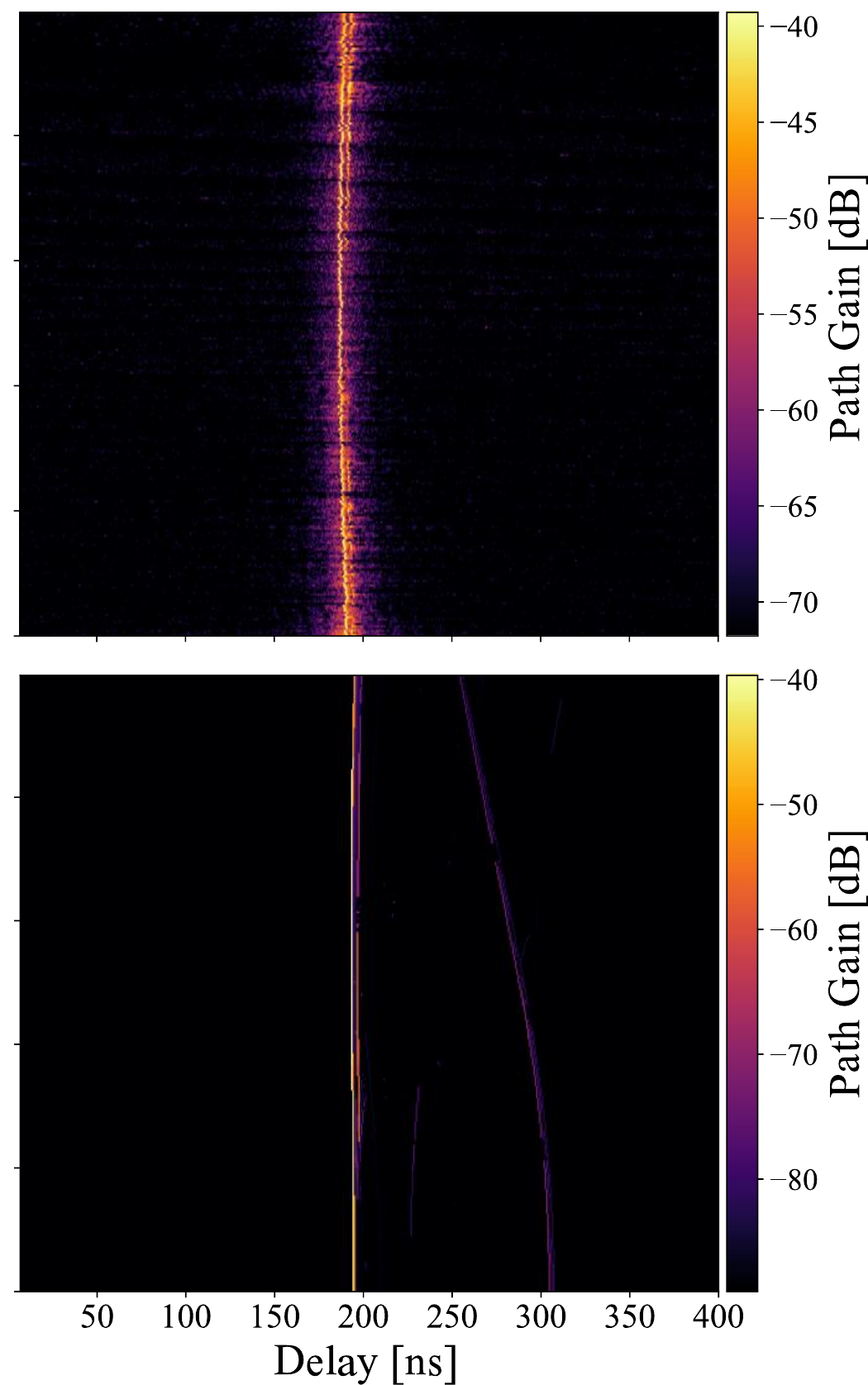}
    \caption{}
    \label{fig:C6_bi}
  \end{subfigure}
  \caption{Scenario~C: PDPs for mono-static (a) and bi-static (b). Top: measurement; bottom: simulation.}
\end{figure}

Fig.~\ref{fig:C6_mono}--\ref{fig:C6_bi} report the Power Delay Profiles (PDPs) for Scenario~C, where the UE moves on a vehicle platform. In the mono-static setup (Fig.~\ref{fig:C6_mono}), both measurement and simulation exhibit multiple time-varying MPCs. These trajectories arise from reflections off nearby objects excited by the UE motion. The RF-DT not only reproduces the delays of these MPCs but also their temporal evolution, demonstrating that the inclusion of diffuse scattering and mobility-aware ray updates is essential to match real-world dynamics.  

In the bi-static setup (Fig.~\ref{fig:C6_bi}), the dominant path remains nearly constant in delay, consistent with the fixed BS--UE geometry. Secondary MPCs linked to environmental objects (e.g., facades, posts) appear at slightly varying delays, again well reproduced by the simulation. This validates the RF-DT’s ability to capture the mixed static/dynamic structure typical of BS-to-UE sensing.

These results highlight that the RF-DT accurately captures mobility-induced variations, not only in terms of path delays, but also in their temporal evolution, which is essential for modeling realistic UE-side sensing scenarios.

\subsection{Doppler Validation}
Beyond delay consistency, it is necessary to confirm whether the RF-DT captures the phase evolution of multipath components within short observation windows, as modeled in~\eqref{eq:doppler}. To this end, Doppler-delay spectra is computed using $N=128$ chirps in \eqref{eq:t_window}, corresponding to a temporal window of $T_w\!\approx\!16$\,ms.

\begin{figure}[t!]
  \centering
  \begin{subfigure}[b]{0.24\textwidth}
    \includegraphics[clip=true,trim=2cm 1cm 6cm 3cm,width=\linewidth]{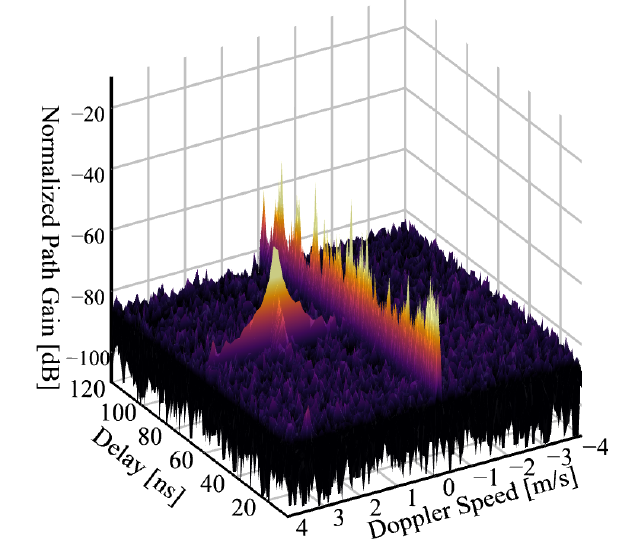}
    \caption{}
    \label{fig:B1_doppler_meas}
  \end{subfigure}\hfill
  \begin{subfigure}[b]{0.24\textwidth}
    \includegraphics[clip=true,trim=2cm 1cm 6cm 3cm,width=\linewidth]{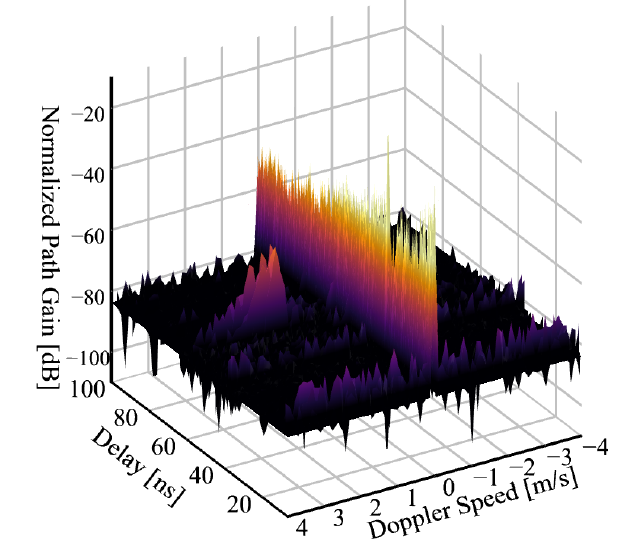}
    \caption{}
    \label{fig:B1_doppler_sim}
  \end{subfigure}  
  \caption{Scenario~B: Doppler-delay spectra from measurement (a) and simulation (b).}
\end{figure}

Fig.~\ref{fig:B1_doppler_meas}--\ref{fig:B1_doppler_sim} illustrate the results for Scenario~B. In both measurement and simulation, a strong ridge centered at zero Doppler is visible, reflecting the contribution of static scatterers such as walls and lamp posts. More importantly, both spectra show a distinct moving component at approximately $2.1$\,m/s and $62$\,ns delay, which corresponds to the trajectory of the vehicle. The ability of the RF-DT to reproduce this peak demonstrates that the simulator correctly maps object motions into Doppler shifts, as predicted by the theoretical model of~\eqref{eq:doppler}. This capability is fundamental in ISAC settings, since Doppler diversity allows the system to discriminate moving targets from static clutter and to directly infer relative velocity.

The close match between the simulated and measured Doppler-delay spectra indicates that the RF-DT preserves both amplitude and phase evolution of dynamic multipath, making it suitable for velocity-aware ISAC algorithm development.

\section{Conclusions}

This work has demonstrated that a calibrated RF-DT based on high-resolution ray tracing can accurately reproduce dynamic ISAC channels in urban microcell environments. Comparisons with measurements across ISAC Scenarios~B and~C, and under both mono-static and bi-static modes, showed that the simulator consistently captured multipath evolution, delay variations from moving scatterers, and Doppler shifts. The close agreement between the simulated and measured power–delay profiles and Doppler–delay spectra confirms that the proposed framework provides a phase-aware and physically consistent representation of real-world propagation.  

Beyond validation, these results highlight the conditions required for reliable RF-DT modeling: accurate 3D scene geometry, material-dependent scattering, realistic antenna patterns, and explicit handling of diffuse and dynamic contributions. By reproducing both static clutter and moving-target signatures, the RF-DT serves as a scalable surrogate for channel measurements, supporting the design of ISAC algorithms, benchmarking, and large-scale scenario exploration. In addition, the framework can be leveraged to generate large and diverse datasets, enabling the training and evaluation of AI- and ML-based methods for sensing, communication, and joint processing. As such, the study bridges the gap between measurement-based characterization and simulation-based prototyping, facilitating robust sensing functionalities in future 6G networks.




\bibliographystyle{IEEEtran}
\bibliography{ref}

\end{document}